\documentstyle[prl,aps,graphicx]{revtex}

\begin{document}
\draft
\twocolumn
\title{Isotope Separation and Solar Neutrino Experiments}
\author{Zakir F. Seidov
 \\
{\it {Department of Physics, Ben-Gurion University of the Negev,
Beer-Sheva 84105, Israel}}}
\maketitle 
\begin{abstract}
Isotopic enrichment of target chemical elements in NSE is 
discussed as means for increasing validity of experimental 
results.
\end{abstract}
{\em{Keywords:}} Solar neutrino experiments; isotope separation.
$$  $$ 
        In spite of  great recent theoretical efforts (see [1] and
     references  therein), the problem of solar neutrinos is still with 
     us: the gap between experimental values and theoretical estimations 
     of solar neutrino fluxes on the Earth is too large as compared with 
the declared uncertainties of both theory and experiment[2]. 
   
     I am not going here to argue with existing (often really brilliant)
     theoretical attempts to find way out in the situation but rather I 
     would like to speculate somewhat about one (even looking fantastic 
     at present) possibility of the drastical improvement at the
     $experimental$ side of the problem. The point is the isotope 
     composition of target elements in Solar Neutrino Experiments (NSE) 
     and problem of improvement of this composition.
      
     Table 1 presents some data on isotope composition of the chemical
     elements suggested by various authors as target in SNE's (see
     comprehensive Compendium [3] and references therein). 
       In each row, first given is $active$ isotope taking part in
neutrino 
       detection reaction, with the relative cosmic abundance of this
       isotope, taking (absolute) cosmic abundance of $^{12}C$ equal to 
       $10^{16}$ (all data on composition are taken from [4]); second 
       shown is $passive$ isotope with its abundance;  $x$ is relative
       abundance of active isotope.
   
       From Table 1 one sees that the "best" is $^{127}I$ (with 100$\%$
relative
       abundance of active isotope) and the "worst" is $^{37}Cl$ with only
       24.2$\%$ of useful isotopes in natural mixture of $Cl$. (Curiously
       enough, $Chlorine$ is characterized in otherwise excellent
review [5]
       as particularly convenient target element having such a $good$ 
       relative abundance.) 
 
       And there is some irony that namely $Chlorine$ is used in the 
       $first$ (and may be $best$) NSE. In huge tank containing
610 tons of
       $perchloretylene$ (dry-cleaning fluid) (or $tetrachloroethene$
$C_2Cl_4$ 
       [127-18-4] by "official" chemical classification) only one-fourth 
       of the whole detector works $hard\  and\  honestly$  while
three-forth
       is only presented but not involved! 
 
       If one would have (in ideal) 100$\%$ pure $^{37} Cl$ and 100$\%$
pure
$^{35}Cl$ in the form of $C_2Cl_4$ both in quality, say, 146 tons, then it
will 
       be crucial in solving experimentally the Problem of Solar Neutrinos 
       (at least in this $radiochemical$ experiment). 

The classical 
       astronomical observation method $star\--background\--star$ (or 
       $on\--off\--on$) will allow to increase enormously the validity of 
       experiment results. 

       The great disadvantage of most (if not all) SNE's is lacking 
       possibility of $switching\  off$  the source to measure the
$background$.\\ 
       The availability of $active$ and $dull$ targets with otherwise
identical 
       characteristics allows - as it is well known - to measure the fluxes 
       which may be even much $smaller\  than\  background$! By the way,
the  volume of target matter may be in this case much smaller than in
       current SNE's.
 
       That is O.K., but what may be done in reality (not in ideal)? -
       quite rightly may ask reader.\\
       That is real question and we forward it to specialists in
       isotope separation [6,7]. What we may do $now\  and\  here$ 
       is only to mention that relative mass difference  $\delta M / M$,
which
       defines largely the elementary separation factor $q_0$ of the all 
       separation methods [6,7], in our case is much larger that for
"sadly famous" $UF_3$. Just as example: for gaseous diffusion $(q_0 -1)
       \propto (\delta M/M)^{1/2}$.
 
       And if one will start to deal with this problem one important
       thing is that - as we guess - there is no need to build special
       plants or installations, as such things are quite in common
       in many countries. Certainly the isotopic-dependant 
       characteristics of $Cl_2$ or $C_2Cl_4$ (the latter being even the
subject 
       of the whole recent book [8]) relevant to isotope separation (see 
       e.g. [9]) may be serve as goal of an international collaboration.  
       
       "$Today\  isotope\  separation\  is\  done\  on\  an\
industrial\\ 
\ scale\  only\  for\ 
       production\  of\  ^{235} U\  and\  heavy\  water$" [6].\\
 May be in near future
       the situation will be drastically changed and $industrial\  scale\ 
       isotope\  separation$ will serve to solve the pure scientific
problem - is it so that thermo-nuclear reactions proceed not only in bombs
       but also in debris of the Sun giving the flux of $10^{11}$ 
neutrinos       per second per cm squared of Earth surface?\\
I am very grateful to J. Bahcall,
I. Dostrovsky, D. Eichler, K. Lande, J. Learned
for comments and discussions.

    
   
\begin{table}
\caption{Some isotopic data on proposed targets in SNE's}
\begin{tabular}{lcccccc}
       Element& Acting&Cosmic&Passive&  Cosm.&  Rel.&  $\delta M/M$\\
          &isotope & abund&  isotope&  abund.& abund., $x$& \\       
\tableline
                                                    
       Lithium &   $^7 Li$ & 45.8 & $^6 Li$ &   3.67& .926&  1/6\\
       Chlorine & $^{37}Cl$ &  1390 & $^{35}Cl$ & 4310& .242 & 2/35\\
       Gallium   &$^{71} Ga$ &19.2  & $^{69}Ga$&29.0&.398&2/69\\
       Bromium   &$^{81} Br$&6.68 &  $^{79}Br$& 6.82& .495& 2/79\\
       Rubidium&$^{87} Rb$ &1.72 &$^{85} Rb$ &4.16 &.293& 2/85\\
       Indium & $^{115} In$& .181&$^{113}In$& .008&.958&2/113\\
       Iodium&$^{127}I$& 1.09& -& -& 1.0& -\\
\end{tabular}
\end{table}
   

\begin{references}
\item[  [1]] 
 Bahcall J.N., Basu S. and Pinsonneault M.P,  astro-ph/9805135;  
          see also hep-ph/9807525, hep-ph/9807216 and
         astro-ph/9805135, astro-ph/9805121.
\item[  [2]]  I should note that along with
a number of more or less
"concordant" SSM's (see references in BP98), 
there is also the somewhat "dissident"
SSM DS98: A. Dar, G. Shaviv, astro-ph/9808098. In this SSM, situation
with
reconciling experimental results and predictions is not so tense.
However I am not in position to judge "who is right, who -  not".
\item[  [3]] 
Bahcall J.N. $Solar\  neutrinos$, Addison-Wesley N.Y. 1994.
\item[  [4]]
Lang K.R. $Astrophysical\  formulae$, Springer N.Y. 1980.
\item[  [5]] 
       Bahcall J.N. $Neutrino\  astrophysics$, CUP N.Y. 1989.
\item[  [6]] 
       Villani S. $Isotope\  Separation$. Am. Nucl. Soc. N.Y. 1972.
\item[  [7]] 
        $Radiochemical\  Separation\  Methods$. Eds. Braun T. and Bujdoso
          Elsevier Amsterdam 1975.
\item[  [8]] 
        $Tetrachlorethylene$. GDCh (Wiss Verlagages) Stuttgart 1996.
\item[  [9]] 
       $Isotope\  Effects\  in\  Chemical\  Reactions$. Eds: Collins C.J.
and
          Bowman N.S. VNR N.Y. 1970.  
  \end{references}
\end{document}